\documentclass[nohyper,11pt,letterpaper]{JHEP3}
\usepackage{graphicx}

\DeclareSymbolFont{AMSa}{U}{msa}{m}{n}
\DeclareSymbolFont{AMSb}{U}{msb}{m}{n}
\let\Box\relax
\DeclareMathSymbol{\Box}{\mathord}{AMSa}{"03}

\def \eqn#1#2{\begin{equation}#2\label{#1}\end{equation}}

\def\gev{\mbox{GeV}}

\title{Perturbation Growth in Anisotropic Cosmologies}

\author{E. Dimastrogiovanni \\  Dipartimento di Fisica ``Galileo Galilei"\\  Universit\`{a} di Padova  and INFN Sezione di Padova\\via F. Marzolo 8 \\ I-35131 Padova, Italy \\E-mail: {\tt  dimastro@pd.infn.it}}

\author{W. Fischler, S.Paban\\
Department of Physics \\ University of Texas, Austin,
TX 78712\\ E-mail: {\tt fischler, paban@zippy.ph.utexas.edu}}

\abstract{We study the growth of perturbations in an expanding Bianchi type-I metric that evolves according to an energy density that includes dust and a cosmological constant. Assuming an epoch where the cosmological constant is subdominant, we find that, for a reasonably large set of initial conditions, the metric fluctuations grow fast enough to make the metric inhomogeneous before the cosmological constant becomes the dominant form of energy. We have examined values for the cosmological constant that are in the interval $10^{10} \mbox{GeV}$ to $10^{16} \mbox{GeV}$.}

\received{???????? ?st, 2000} \accepted{???????? ?th, 1998}
\preprint{UTTG-02-08}

\begin{document}



\section{\bf Introduction}

Inflation offers so far the best description for the  power spectrum of density fluctuations with its slight deviation from scale invariance.  The claims that inflation explains the ``homogeneity and flatness" are more questionable. There is an extensive literature on trying to ascertain under which conditions for the initial metric and the energy-density will a space develop a period of inflation \cite{Gibbons:1977mu}-\cite{Berera:2000xz}. Although there is wide spread belief that space-times with a cosmological constant isotropize at late times, there is no complete formal proof \cite{Rendall:2004kh}.  One result on which there seems to be general agreement is Wald's proof  \cite{Wald:1983ky} that all expanding anisotropic but homogenous  models in the presence of a positive cosmological constant, with the exception of some Bianchi-IX models,  evolve toward a de Sitter solution. In the case of the Bianchi-IX this result is also true provided the cosmological constant is sufficiently large compared to spatial curvature terms. Recent work has also been directed toward understanding the possible experimental signatures of an initial universe that is homogeneous but anisotropic \cite{Pereira:2007yy} - \cite{Gumrukcuoglu:2007bx}. 
  
In this work we consider an initially expanding type-I Bianchi metric that evolves according to an energy density not yet dominated by  a cosmological constant.  This metric is one of the special cases studied by Wald \cite{Wald:1983ky}. According to his work, if the metric maintains its homogeneity until the cosmological constant becomes the dominant form of energy, it is expected to  asymptote to a de Sitter cosmology.  The fluctuations of this metric, however, will  make it evolve from an anisotropic but homogeneous metric into an anisotropic and inhomogenous one. The question we seek to answer is: will this growth be fast enough as to invalidate the assumption of homogeneity before the cosmological constant becomes the dominant form of energy? We will see that this is indeed the case for a reasonably large set of initial conditions, when we assume for the cosmological constant to be in the interval between $10^{10} \gev$ and $10^{16} \gev$. 

In the next section we will summarize the known results about the time evolution of the type-I Bianchi models and its fluctuations, and in section 3  we will present our results.

{\section{\bf Evolution of Type-I Bianchi and its fluctuations} }

The type-I Bianchi metric is of the form

\eqn{metric}{ {ds}^2= -dt^2 + e^{2 \alpha} \, (e^{ 2 {\beta}})_{ij} \, dx^{i}  dx^{j} } 
where $\alpha=\alpha(t)$  and characterizes the volume expansion, and $\beta_{ij}=\beta_{ij}(t)$ is a $3 \times 3$ diagonal traceless metric that describes the anisotropy,

\eqn{beta}{(e^{2 \beta})_{ij}= \delta_{ij} \, e^{ 2 \beta_i}  \,   \,\,\,\,\,\,\,\,\,\,\,\,\,\,\,\,\,   \sum_{i=1}^{3} \beta_i=0}
The volume factor is only a function of $\alpha$, $\sqrt{-g}= e^{3 \alpha}$.
The equations of motion are described in  Appendix-A, in this section we will only present the solutions for the dust background, that is for an energy momentum tensor whose only non-vanishing component is $T_{00}= w_0 \, e^{-3 \alpha}$.  In this model we have made the simplifying assumption that the unperturbed energy-momentum tensor is isotropic, placing all the anisotropy  on the initial metric configuration. We have chosen this  energy-momentum  to take advantage of the existing analytical solution.  In addition, this choice of the energy-momentum tensor will not exacerbate the anisotropy, as the solution below shows. Therefore,  if non-trivial results are obtained for this example, we expect this behavior to generalize to more anisotropic forms of the unperturbed energy momentum tensor. 
\eqn{betasol}{\beta_{ij}= b_{ij} \, u + c_{ij} }
is a solution provided we identify

\eqn{udef}{ du= e^{-3 \alpha} \, dt}
and the matrices  $b_{ij}$ and $c_{ij}$ are traceless, diagonal and constant. The functions in the metric evolve as:

\begin{eqnarray}
e^{3 \alpha}  & = &  \frac{ 3 \, w_0 }{4} \, t ( t + t_b) \nonumber \\
\label{alpha-beta} \\
e^{\beta_i}  & = &  \left( \frac{t}{t + t_b} \right)^{s_i/3} \nonumber 
\end{eqnarray}
In this solution,  $b^2= \frac{1}{6} b_{ij}b_{ij}$, $t_b=4 b / {w_0}$,  $s_i=b_{ii}/b$ and   the $c_{ij}$ have been set equal  to zero. The $s_i$ determine the shape of the anisotropy and satisfy the constraints:

\eqn{s-constraints}{\sum_{i=1}^{3}  s_{i}^2 = 6, \,\,\,\,\,\,\,\,\,\,\,\,\,\, \sum_{i=1}^{3}  s_{i}= 0}

From these solutions we can observe that at very late times $t >>t_b$ the effect of the anisotropy is washed out and the metric evolves as an isotropic FRW cosmology in the presence of dust. It is for $t<<t_b$ that the model shows its maximum anisotropic behavior. Indeed,  although the overall volume factor, $\sqrt{-g}$, increases, this is not necessarily the case for the direction-dependent scale factors.   Labeling the $s_i$ exponents in increasing order, the constraints (\ref{s-constraints}) require that
\eqn{s}{ -2 \leq s_1 \leq -1 \leq s_2 \leq 1 \leq s_3 \leq 2 }
(A note of caution: in the remaining of the paper we might use a different order for these indices, and $s_3$ can be a negative number.) The different components of the metric in the limit $t<<t_b$, evolve as:

\eqn{metric-components}{{a_i}^2  \equiv g_{ii} =e^{ 2 \alpha + 2 \beta_i}  \sim t^{2 (1+s_i)/3}}
With the exception of the point $(s_1,s_2,s_3)=(-1,-1,2)$ the other solutions to (\ref{s-constraints}) give cosmologies on which, at least, one direction is contracting.

 A complete Bardeen-type analysis of the perturbations to this metric and identification of gauge invariant quantities has been  recently done by Pereira, Pitrou and Uzan \cite{Pereira:2007yy}. Because our work was started before their work was published,  we use an earlier work of Perko, Matzner and Shepley (PMS) \cite{Perko:1972cs}, where the equations are solved  in  the synchronous gauge instead. Since the synchronous gauge leaves a remaining gauge invariance,  care will have to be taken to assure that the results that we use are not a gauge artifact but actually reflect a physical behavior. In our case, since we restrict our study to dust, the synchronous gauge is co-moving and the gauge is completely fixed, removing any source for concern on this issue. Earlier work  by Noh \cite{Noh:1995se}( see also \cite{Tomita:1985me}) who repeated Perko, Matner and Shepley computation in the co-moving gauge, showed agreement  in the quantities  used in this work.

In the synchronous gauge, the perturbed metric  can be written as:
\begin{equation}
ds^2=-\frac{1}{\gamma}d\tau^2+e^{2 \alpha}(e^{2 \beta})_{ik}\left(\delta^{k}_{j}+h_{j}^{k}(\tau,\vec{x})\right)dx^{i}dx^{j}
\end{equation}  
The energy momentum tensor for a perfect fluid in its rest frame is

\begin{equation}\label{11}
T_{\mu\nu}=pg_{\mu\nu}+(p+\omega)u_{\mu}u_{\nu},
\end{equation} 
where $u_{\mu}=(-\frac{1}{\sqrt{\gamma}},0)$, $u^{\mu}=(-\sqrt{\gamma},0)$. The perturbed density, pressure and fluid velocity are written as

\begin{eqnarray}\label{12}
\omega' & = &\omega+\delta \omega \nonumber \\
p' & = & p+\delta p  \\
u'^{\mu} & = & u^{\mu}+\delta u^{\mu} \nonumber
\end{eqnarray}
With our gauge choice $\delta u^{0}=\delta u_{0}=0$, while $\delta u^{i}\neq 0$.\\

The details of the perturbations analysis can be found in Appendix B. The metric tensor decomposes into two scalar, two vector and two tensor degrees of freedom. Our restriction to  a matter fluid, although maybe not as realistic as a radiation fluid, greatly simplifies the calculations since it is possible to operate in a comoving (in addition to synchronous) gauge. Also, the wave vector of the Fourier components of the modes is for simplicity fixed along one of the spacial directions (specifically, the third direction, $\vec{k}=(0,0,k_3)$). With these assumptions, one of the tensor modes is free and the vector modes  decouple from other perturbations. The equations of interest to our analysis involve a scalar, the density contrast (the other scalar modes depend on it), and the remaining tensor degree of freedom, which are coupled to each other

\begin{eqnarray}\label{1}
& &{\left( \frac{d^2 \delta}{dt^2} \right)\{(FK^2t^2)
(t+t_b)^2+\frac{1}{4}s_3^2t_b^2} \}-
\frac{2}{3t}\frac{1}{(t+t_b)}\left((t+t_b)^2(FK^2t^2)-\frac{3}{4}t_b^2s_3^2\right)\delta \nonumber \\  \nonumber \\
&+ & \left[\frac{(t+t_b)}{t}(FK^2t^2)\left(\frac{2s_3}{3}t_b+\frac{2}{3}(2t+t_b)\right)+
t_b^2s_3^2\frac{(2t+t_b)}{2t(t+t_b)}\right]\frac{d \delta}{dt} \nonumber \\
&= &\frac{1}{3}(s_1-s_2)(FK^2t^2)
\left[\frac{1}{2}s_3\left(\frac{t_b}{t}\right)^2\eta+
\frac{t_b}{t} (t+t_b) \frac{d \eta}{dt}\right] 
\end{eqnarray}

\eqn{2}{\frac{d^{2}\eta}{dt^2}+\frac{2t+t_b}{t(t+t_b)}\frac{d \eta}{dt}+\\
\eta K^2F=\frac{1}{3}(s_1-s_2)\frac{t_b}{t(t+t_b)}\frac{d \delta}{dt}}
where $\delta$ and $\eta$ are respectively the Fourier components of the density constrast and of the tensor mode (the Fourier index $k$ has been suppressed), $$F\equiv \left(\frac{1}{t(t+t_b)}\right)^{\frac{2}{3}}\left(\frac{t+t_b}{t}\right)^{\frac{2s_3}{3}} \hspace{4ex} \mbox{and} \hspace{4ex}  K\equiv k_3 \left(\frac{4}{3 \omega_0}\right)^\frac{1}{3}$$
Depending on the value of $s_3$ we can distinguish two cases:

\begin{description}

\item[{(I)}]  $s_{3}= \pm 2$ (i.e. in the axially symmetric case $s_{1}=s_{2}$), the density contrast and the tensor mode decouple and the solutions are

\eqn{dni}{ \delta = \left\{ \begin{array}{ll} 
\frac{A}{t}+B\left[\frac{5t_b}{t+t_b}+3K^2\frac{(t+t_b)^{\frac{5}{3}}}{t}\right] &  \,\,\,\,\,\,\, \mbox{for} \,\,\,\,\,\, s_3=2 \\   \\ 
\frac{A}{t+t_b}+B\left[-\frac{5t_b}{t}+3K^2\frac{t^{\frac{5}{3}}}{t+t_b}\right] & \,\,\,\,\,\,\, \mbox{for} \,\,\,\,\,\, s_3=-2
\end{array}\right. }

where A and B are constants of integration.\\

\item[{(II)}] $s_{3}\neq \pm 2$, analytic solutions can be found in the long ($FK^2t^2 \ll 1$) and short ($FK^2t^2 \gg 1$) wavelength limits. For long wavelengths we have

\eqn{5}{
\delta=\frac{A}{t}+B+Ct^{\frac{2(2-s_3)}{3}}\left(1+D \ln\frac{t}{t_b}\right)}

and for short wavelengths

\eqn{31}{\delta=At^{\frac{(1-2s_3)}{3}}+B}
These solutions are derived in Appendix B.
\end{description}

\section{\bf Study of Perturbations}

In the previous section we derived the time dependence of the matter perturbations. In this section, we will introduce a useful parametrization  for analyzing their growth.
We will not be interested in studying the axially symmetric cases $s_3=\pm 2$ for two different reasons. When   $s_3=+2$    the perturbation doesn't grow in time (\ref{dni})  and hence it can never become important.  For  $s_3=-2$ the growth is wavelength dependent and  would require a separate investigation.
In the limit $t\ll t_b$ and $s_3\not=\pm 2$, for the perturbation that is outside the horizon, the fastest growth is given by:

$$ \delta\sim t^{\frac{2(2-s_3)}{3}}$$
while inside the horizon

$$ \delta\sim t^{\frac{(1-2s_3)}{3}} $$ 
So as long as $ -2<s_3<\frac{1}{2}$ the perturbations will grow both inside and outside the horizon. Let's introduce the following parametrization

\begin{eqnarray}\label{7}
\delta & = & 10^{-n}\left(\frac{t}{t_P}\right)^s= \delta_P \left(\frac{t}{t_P}\right)^s \\
t_b & = & 10^{n_b}t_{P} 
\end{eqnarray} 
where $t_P=5 \times 10^{-44} s$ is the Planck time, $n$ and $n_b$ are positive integers and $s$ stands for the power $s_>\equiv\frac{2(2-s_3)}{3}$ outside the horizon or for $s_<\equiv\frac{(1-2s_3)}{3}$ inside the horizon.\\
The time $t_1$ at which the perturbation becomes of order one is

\begin{equation}
\delta|_{t=t_1}=1 \Longrightarrow t_1=t_b10^{-n_{b}+\frac{n}{s}}
\end{equation}

For consistency, we shall  impose the condition $t_1\ll t_b$ since the perturbation analysis carried out so far is for times well before the anisotropy-FRW transition. It then follows that $n$ and $n_b$ are constrained by 

\begin{equation}\label{6}
n_b>\frac{n}{s}
\end{equation}

The total energy density for a pressureless fluid at $t\ll t_b$ is $\omega \sim \frac{M_P^2}{tt_b}$. From the preceeding equations it follows that

\begin{equation}
\omega|_{t=t_1}\equiv \omega_1 \sim M_P^4 10^{-(n_b+\frac{n}{s})}
\end{equation}

Let's parametrize the scale of inflation $\Lambda=M_P 10^{-q}$, where $q$ is a positive integer. For the matter perturbations to reach order one well before the universe starts to inflate,  ${\omega_1}^{\frac{1}{4}} \gg \Lambda$, which translates into the constraint $n_b+\frac{n}{s} < 4q$. This last equation, combined with (\ref{6}) produces the condition $\frac{n}{s} < 2q$, which will be later employed in order to derive a possible range of values for $t_1$, $t_b$, $\omega_1$.\\

\subsection{Growth of perturbations at different wavelengths}

Although the growth of the scale factor, for  $t \ll t_b$,  is different for different directions, 

\[ a_i(t)= e^{\alpha+\beta_i }= \left\{ \begin{array}{ll} \left(\frac{3 \omega_{0}}{4}\right)^{\frac{1}{3}}t^{\frac{1+s_{i}}{3}}t_b^{\frac{1-s_{i}}{3}}  &   \hspace{6ex} t \ll t_b  \\
\left(\frac{3 \omega_{0}}{4}\right)^{\frac{1}{3}}t^{\frac{2}{3}} &  \hspace{6ex} t \gg t_b \end{array}\right.   \]
the horizon scale has the same growth for all of them.

\[ H_i(t)= \left\{ \begin{array}{ll} \left(\frac{1+s_i}{3}\right)\frac{1}{t}  &     \hspace{6ex} t \ll t_b  \\   \frac{2}{3t}  &    \hspace{6ex}  t \gg t_b  \end{array}\right.   \]
As already seen in the previous section, and in analogy with the isotropic case, the perturbation growth depends on the size of the physical wavelength in regards to the horizon scale. Indeed, the critical quantity $FK^2t^2$ is nothing but

$$ F K^2 t^2  = \mbox{constant} \left(  \frac{k_3}{a_3 H} \right)^2 $$
where $H \sim 1/t$. The product $a_3 H$ evolves with time as:
\[ \ a_3 H (t) \sim \left\{ \begin{array}{ll} t^{\frac{s_{3}-2}{3}}t_b^{\frac{1-s_{3}}{3}}  &  \hspace{6ex} t \ll t_b  \\ t^{\frac{-1}{3}}  &  \hspace{6ex} t \gg t_b  \end{array}\right.   \]
Thus   wavelengths that are inside the horizon at early times (close to the Plank scale) will remain inside the horizon at later times. 
The wavelength $\lambda$ decreases as a function of time until horizon crossing for $-2<s_3< -1$  and increases if $-1<s_3< \frac{1}{2}$.

It is important to search for constraints to the parameters involved in the perturbation analysis.
Let's begin from the time $t_b$ of transition between anisotropy-dominated and FRW-like universe. The universe appears isotropic and homogeneous up to the highest accessible redshifts, therefore an upper bound for $t_b$ would be around $13.7 \times 10^9  yr$. The lower bound would obviously be reprensented by the Plank time.
The scale of inflation has been constrained from above to $10^{16} \mbox{GeV}$ by the gravitational wave experiments \cite{Lidsey:1994xv}. This limit translates into the condition  $ 3 \leq q $.  

No lower observational bound on the initial values of the matter perturbations can be determined since the perturbations would grow to order one and the outcome of non linear theory would be independent of initial conditions.

Finally it is important to remember that the expressions derived for the matter perturbations are valid in two special limits, short and long wavelengths, and the values of $q$ have to be in the intervals derived in this section.
The lower bound on $q$ and upper bound on $t_b$  can be used to further constrain $n$ and $n_b$, as follows 

\begin{eqnarray}
q=3 \Longrightarrow n<6s, \hspace{3ex} \frac{n}{s}<n_b<12-\frac{n}{s},\\
t_b < 10^{60} t_P \Longrightarrow n_{b} < 60.
\end{eqnarray}

In the Table \ref{table1} the parameters have been calculated under the condition that the (inside the horizon) matter perturbations reach order one before the time corresponding to the earliest possible beginning for inflation. 
\\ 
\begin{table}[h]
\caption{ Range of variation of parameters for $s_3=- \sqrt{3}$, q=3} \vspace{2ex}
\begin{tabular}{|c|c|c|c|c|c|}\hline
n & $n_{b}$ & $\delta_P$ & $t_1$ (in units of $t_P$) & $\left(\omega_{1}\right)^{\frac{1}{4}}$ (in units of $M_P$) & $t_b$ (in units of $t_P$) \\
\hline\hline
8 & 5.3 to 6.6 & $10^{-8}$ & $2(10^{5})$ & $10^{-3}$ to $2(10^{-3})$ & $t_1$  to $4(10^{6})$  \\
\hline
6 & 4.0 to 8.0 & $10^{-6}$ & $10^{4}$  & $10^{-3}$ to $10^{-2}$  & $t_1$  to $10^{8}$\\
\hline
4&2.7 to 9.3 & $10^{-4}$&$5(10^{2})$&$10^{-3}$ to $4(10^{-2})$& $t_1$  to $2(10^{9})$\\
\hline
2& 1.3 to 10.6&$10^{-2}$&20&$10^{-3}$ to 0.2&$t_1$  to $4(10^{10})$\\
\hline
$10^{-1}$&0.06 to 11.9&0.8&1&$10^{-3}$ to 0.9&$t_1$  to $7(10^{11})$\\
\hline
\end{tabular}
\label{table1}
\end{table}
\\
\\
As expected from the definition  of $n$ (\ref{7}), the perturbations reach order one at a smaller and smaller time as $n$ is given smaller and smaller values. Also as $n$ approaches zero, the range of allowed values for $n_b$ becames larger and larger, and consequently the same happens for the range of possible values of $t_b$. 
If the upper bound on the inflationary energy scale was pushed down by a few orders of magnitude, the range of possible values for $n$ and $n_b$ would be broader and the same conclusion would apply to $t_1$ and $t_b$ as shown in  Table  \ref{table2}
\\
\begin{table}[h]
\caption{ Range of variation of parameters for $s_3=- \sqrt{3}$, q=9} \vspace{2ex}
\begin{tabular}{|c|c|c|c|c|c|}\hline
n & $n_{b}$ & $\delta_P$ & $t_1$ (in units of $t_P$) & $\left(\omega_{1}\right)^{\frac{1}{4}}$ (in units of $M_P$) & $t_b$ (in units of $t_P$) \\
\hline\hline
25 & 16.8 to 19.2 & $10^{-25}$ & $6(10^{16})$ & $10^{-9}$ to $4(10^{-9})$ & $t_1$  to $10^{19}$  \\
\hline
20 & 13.4 to 22.6 & $10^{-20}$ & $2(10^{13})$  & $10^{-9}$ to $2(10^{-7})$  & $t_1$  to $4(10^{22})$\\
\hline
15&10.1 to 25.9 & $10^{-15}$&$10^{10}$&$10^{-9}$ to $9(10^{-6})$& $t_1$  to $8(10^{25})$\\
\hline
10& 6.7 to 29.3&$10^{-10}$&$5(10^{6})$ &$10^{-9}$ to$4(10^{-4})$ &$t_1$  to $2(10^{29})$\\
\hline
5&3.3 to 32.6&$10^{-5}$& $2(10^{3})$&$10^{-9}$ to $2(10^{-2})$&$t_1$  to $4(10^{32})$\\
\hline
1&0.7 to 35.5&$10^{-1}$&5&$10^{-9}$ to 0.4&$t_1$  to $2(10^{35})$\\
\hline
\end{tabular} 
\label{table2}
\end{table}
\\

\subsection{Final results} 

The results of the previous tables correspond to wavelengths that start inside the horizon and  remain there, but there are other options. Depending on the value of $s_3$ we can distinguish between 
\begin{description}

\item[{(a)}] $-2<s_3<-1$, in this region the physical wavelength decreases with time as long as $t \ll t_b$.
\item[{(b)}] $-1<s_3<\frac{1}{2}$, in this region the physical wavelength increases with time although not as fast as $H^{-1}(t)$
\end{description}
In these two regions it is possible to distinguish between the following two behaviors:

\begin{itemize}
\item the wavelength is outside the horizon at $t=t_i$ (initial time), which will occur for comoving wavelengths

$$ \lambda_0 > \lambda_{0}^{H}\equiv\frac{3}{|1+s_3|}\left(\frac{4}{3\omega_{0}}\right)^{\frac{1}{3}}t_i^{\frac{2-s_3}{3}}t_b^{\frac{s_3-1}{3}}$$

Then, the perturbation crosses inside the horizon at some time $t_{HC}$. Although $t_{HC}$ can be both bigger of smaller than $t_b$, only the case when is smaller will be of interest to us since we have been making the assumption $t \ll t_b$.
For a comoving wavelength $\lambda_0$,  the time of horizon crossing $t_{HC}$ is

$$t_{HC}=\left[\lambda_{0}\left(\frac{3\omega_{0}}{4}\right)^{\frac{1}{3}}\frac{|1+s_{3}|}{3}t_{b}^{\frac{1-s_{3}}{3}}\right]^{\frac{3}{2-s_{3}}}$$
The constraint $t_{HC} < t_b$ puts an upper bound on the comoving wavelength 

$$ \lambda_0 < \lambda_0^{HH}\equiv \frac{3}{|1-s_{3}|}\left(\frac{4}{3\omega_{0}}\right)^{\frac{1}{3}}t_{b}^{\frac{1}{3}}$$

\item the wavelength is inside the horizon at $t=t_i$, that is $\lambda_0 < \lambda_0^H$.  The perturbation will  never cross outside the horizon  since the slope for $\lambda(t)$ is negative at small times and goes like $t^{\frac{2}{3}}$ at later times, while the slope of $H^{-1}$ grows always like $t$.
\end{itemize}

\subsection{Comparison with FRW Universe}

In this section we will compare the results that we have just obtained with those of an ordinary matte dominated FRW universe. The perturbations both inside and outside the horizon grow as \cite{Lifshitz:1963ps}, \cite{Mukhanov:1990me} 

$$\delta=\delta_i \left(\frac{t}{t_i}\right)^{\frac{2}{3}}$$
while the energy density changes as 

$$\omega=\omega_i \left(\frac{t}{t_i}\right)^{-2}$$
In analogy with the parametrization introduced in (\ref{7}),
let's parametrize the initial perturbations as 

$$\delta=10^{-n} \left(\frac{t}{t_P}\right)^{\frac{2}{3}}$$ and
choose $t_i=10t_P$, in the event that the universe starts out homogeneous and isotropic.\\
The time at which the perturbations become of order one is $t_1=10^{\frac{3n}{2}}t_P$.\\
Summarizing

\begin{eqnarray}
&t_1^{FRW}=10^{\frac{3n}{2}}t_P\\
&t_1^{BI}=10^{\frac{n}{s}}t_P,
\end{eqnarray}
where 

\[ s= \left\{ \begin{array}{r@{\quad:\quad}l} \frac{2(2-s_3)}{3}  & $at large wavelenghts$  \\ \frac{1-2s_3}{3}  & $at small wavelengths$.  \end{array}\right.   \]
For the growth of perturbations for the Bianchi I model to be faster than in the FRW case: $t_1^{BI} < t_1^{FRW}$; this constraint implies that the quantity $s>\frac{2}{3}$, which translates into the following bound for $s_3$.

\[ \left\{ \begin{array}{r@{\quad:\quad}l} s_3<1   & $for large wavelenghts$  \\
s_3<-\frac{1}{2} & $for small wavelengths$.  \end{array}\right.   \]

\begin{table}[h]
\caption{ Comparison between the time that it takes for perturbations to grow to be of order one in the Bianchi type I model and in the FRW model, in the special case where $s_3=- \sqrt{3}$. The quantity $n$ measures the size of the initial perturbations, $n=\frac{2}{3}-\log{ \delta_i}$  for the FRW case, and $n=s- \log {\delta_i}$ in the Bianchi case.} \vspace{2ex}
\begin{tabular}{|c|c|c|}\hline
n & $t_1^{BI}$ (in units of $t_P$) &  $t_1^{FRW}$ (in units of $t_P$) \\
\hline \hline
25 & $6(10^{16})$ & $3(10^{37})$ \\
\hline
20 & $2(10^{13})$ & $10^{30}$ \\
\hline
15& $10^{10}$&$3(10^{22})$\\
\hline
10 &$5(10^{6})$&$10^{15}$ \\
\hline
6& $10^{4}$ & $10^{9}$\\
\hline
4& $5(10^{2})$ & $10^{6}$\\
\hline
2&20&$10^{3}$\\
\hline
\end{tabular}
\end{table}

\section{\bf Conclusions}

The Bianchi type I model (homogeneous and anisotropic) was introduced as a background spacetime for a cosmological model characterized by small density inhomogeneities.
The evolution of these perturbations was investigated for a pressureless cosmic fluid and the equations turned out to be complicated by the presence of a coupling between the density contrast and the tensor perturbation modes. This coupling is absent for special choices of the anisotropy parameters and for isotropic models in general.
Analytic solutions in the limits of small and large wavelength perturbations (compared to the horizon size) were found. The tensor mode that is coupled to the scalar energy mode turned out to be an oscillatory function of time with an amplitude decreasing in time for any possible value of the anisotropy parameter, whereas the density contrast showed a power law growing behaviour in time for a large range of parameter space. The growth is faster in the contracting direction of space compared to the directions that are in expansion for both small and large wavelength limits. These suggestive results motivated us to further analyze the growth of density inhomogeneities, specially,  in view of the possibility that inflation might not occur if the spacetime is not homogenous. For quite a large range of parameters of the theory, the inhomogeneities in the energy density will grow to be of order one at a time where $\Lambda$ is still a subdominant component of the energy density. 
As the period available for the fluctuations to become non-perturbative grows, while the cosmological constant remains subdominant, the restrictions on the parameter space will weaken substantially.

Since our analysis is restricted to perturbations, we cannot follow the evolution of inhomogeneities to the non-perturbative domain. In order to properly study the subsequent evolution of the universe, non-perturbative tools will have to be used. The results presented in this paper raise serious concern to motivate further work to find the fate of the universe in this context.

\section{Acknowledgments}

The work of E. Dimastrogiovanni, W. Fischler and  S. Paban has been
partially supported by the National Science Foundation under Grant No. PHY-0455649.

\section{\bf Appendix A}
The Kasner metric is generally written as

\eqn{kasner}{ds^2=-dt^2+t^{2p_1}\, dx_1^2+t^{2p_2}\, dx_2^2+t^{2p_3}\, dx_3^2}
The coefficients $p_1$, $p_2$, $p_3$ are constants satisfying the relation

\eqn{kasner1}{p_1+p_2+p_3=p_1^2+p_2^2+p_3^2=1}
The three parameters can be equal in pairs in the cases $(-\frac{1}{3}, \frac{2}{3}, \frac{2}{3})$ and $(0, 0, 1)$. In all the other cases they are distinct, one being negative and the other two being positive. Under the assumption $p_{3}<p_{2}<p_{1}$, the allowed ranges follow from (\ref{kasner1}) 

\begin{eqnarray*}
\frac{2}{3}\le &p_{1} \le 1,\\
0\le &p_{2} \le \frac{2}{3},\\
-\frac{1}{3}\le &p_{3} \le 0.
\end{eqnarray*}

The coefficients can be parametrized as 

\begin{eqnarray*}
&p_{1}(u)=\frac{u(1+u)}{1+u+u^{2}},\\
&p_{2}(u)=\frac{1+u}{1+u+u^{2}},\\
&p_{3}(u)=\frac{-u}{1+u+u^{2}},\\
\end{eqnarray*}
where $u>1$ .

The universe described by the metric (\ref{kasner}) is spatially flat and, for any value of the coefficients $p_i$, in expansion since the volume element is $ \sqrt{g^{(3)}} \, d^3x = t \, d^3x $.\\

From (\ref{kasner1}) it is evident that the arguments $p_i$ of the scale factors in the different directions of space cannot have the same sign, which indicates that the universe is expanding in some directions and contracting in some other directions with a law like $L_i^{p}=L_i^{c} \, t^{p_{i}}$,
where $L_i^{p}$ and $L_i^{c}$ are respectively proper and comoving distances.\\

A metric of the kind (\ref{kasner}) can be generalized introducing a time dependence for the coefficients and can be put in the form   

\eqn{bianchiI}{ds^2=-\frac{1}{\gamma}d\tau^{2}+e^{2\alpha}(e^{2\beta})_{ij}dx^{i}dx^{j}}
where $\alpha=\alpha(\tau)$, $\beta_{ij}=\beta(\tau)_{ij}$ is a traceless diagonal $3\times3$ matrix. The function $\gamma=\gamma(t)$ can be eliminated redefining the time as $dt=\gamma^{-\frac{1}{2}}d\tau$.

The equations for this metric, in the presence of an isotropic perfect fluid, were derived for the first time by \cite{Perko:1972cs} and later also by \cite{Noh:1995se}.

\begin{eqnarray}
-\frac{3}{2}\dot{\alpha}\dot{\gamma}-3\gamma\ddot{\alpha}-3\gamma\dot{\alpha}^{2}-\gamma\dot{\beta_{ij}}\dot{\beta_{ij}}& = & \frac{1}{2}(\omega+3p)\\
\frac{1}{2}\dot{\alpha}\dot{\gamma}+\gamma\ddot{\alpha}+3\gamma(\dot{\alpha})^{2}& = & \frac{1}{2}(\omega-p)\\
\dot{\gamma}\dot{\beta_{ij}}+2\gamma\ddot{\beta_{ij}}+6\gamma\dot{\alpha}\dot{\beta_{ij}} & = & 0,
\end{eqnarray}
where the dot indicates the derivative with respect to $\tau$. It is easy to check that 

\eqn{20}{\beta_{ij}=b_{ij}u+c_{ij}  }
solves the field equations if $u$ acts as a new time coordinate 

\begin{equation}
du=\frac{e^{-3\alpha}} {\sqrt{\gamma}}d\tau \label{21}
\end{equation}
with $b_{ij}$ and $c_{ij}$ traceless diagonal constant constant matrices.
The relation (\ref{kasner1}) can be translated into a condition involving the $b_{ij}$.
The first of the conditions (\ref{kasner1}) translates  into the requirement that $\mbox{Tr}(\beta_{ij})=0$, while
the second  condition (\ref{kasner1}) corresponds to

 $$ \mbox{Tr} \left[\delta_{ij}+ 2\frac{d \beta_{ij}}{d \alpha}+\left(\frac{d \beta_{ij}}{d \alpha}\right)^{2}\right]=9$$ which is equivalent to

\begin{equation}
\sum_{i=1,2,3}\left(\frac{ d\beta_{ii}}{d \alpha}\right)^{2}=6.
\end{equation}
Plugging (\ref{20}) and (\ref{21}) into Einstein equations, we have

\begin{eqnarray}
\left(\frac{d\alpha}{du}\right)^2 & = & \frac{1}{3}e^{6\alpha}\omega+b^{2}\\
\frac{d\omega}{du}& = & -3\frac{d\alpha}{du}(\omega+p).\label{22}
\end{eqnarray}
Integrating, for $p= \sigma \omega$,  we have

 $$e^{3(\sigma-1)\alpha/2}=-\sqrt{\frac{\omega_{o}}{3b^{2}}}\sinh\left[\frac{3}{2}b (1-\sigma) \, u\right]$$ 
where $b^{2}\equiv\frac{1}{6}b_{ij}b_{ij}$ and $\omega_0$ is an integration constant. For the special case of pressureless matter $(\sigma=0)$, in terms of the proper time $t$

\begin{equation}
e^{3\alpha}=\left(\frac{3\omega_{o}}{4}\right)t\left(t+4\frac{b}{\omega_{o}}\right).
\end{equation}

\section{Appendix  B}

The perturbed metric is $ds^2=-\frac{1}{\gamma}d\tau^2+e^{2 \alpha}e^{2 \beta}_{ik}\left(\delta^{k}_{j}+h_{j}^{k}(\tau,\vec{x})\right)dx^{i}dx^{j}$.
 
The unperturbed affine connection is\\

\[ \begin{array}{lll}
\Gamma^{0}_{00}=\frac{1}{2}g^{00}\frac{dg_{00}}{dt} & \hspace{5ex}  & \Gamma^{0}_{0i}=0 \\
\Gamma^{k}_{0i}=\frac{1}{2}g^{kl}\frac{dg_{li}}{dt} &  &  \Gamma^{i}_{ki}=0  \\ 
\Gamma^{k}_{00}=0  & &   \Gamma^{0}_{ki}=-\frac{1}{2}g^{00}\frac{dg_{ki}}{dt} \\
\Gamma^{j}_{ki}=0 & & \\
\end{array}  \]
\\
The first order perturbation in the affine connection is\\

\[ \begin{array}{lll}
\delta\Gamma^{0}_{00}=0  & \hspace{3ex} & \delta\Gamma^{0}_{0i}=0\\ 
\delta\Gamma^{k}_{0i}=\frac{1}{2}g^{kj}\frac{\partial{\delta g_{ij}}}{\partial{t}}+\frac{1}{2}\delta g^{kj}\frac{\partial{g_{ij}}}{\partial{t}}&  &\delta\Gamma^{k}_{00}=0 \\
\delta\Gamma^{0}_{ki}=-\frac{1}{2}g^{00}\frac{\partial{\delta g_{ki}}}{\partial{t}}& &\delta\Gamma^{j}_{ki}=\frac{1}{2}g^{jl}(\frac{\partial{\delta g_{il}}}{\partial{x^{k}}}+\frac{\partial{\delta g_{kl}}}{\partial{x^{i}}}-\frac{\partial{\delta g_{ki}}}{\partial{x^{l}}})
\end{array}  \]
The perturbed first order Einstein equations are 

\begin{equation}
\delta R_{\mu\nu}=\delta T_{\mu\nu}-\frac{1}{2}\delta g_{\mu\nu}T-\frac{1}{2}g_{\mu\nu}(\delta g^{\sigma \lambda}T_{\sigma \lambda}+g^{\sigma \lambda}\delta T_{\sigma \lambda}).
\end{equation}
For a perfect cosmic fluid in a comoving frame we have \cite{Perko:1972cs}

\begin{eqnarray} \label{eq}
&- & \frac{1}{2}\dot{\gamma}\dot{h}- \gamma(\ddot{h}+2 \dot{\alpha}\dot{h}+2\dot{\beta_{ij}}\dot{h^{i}_{j}})=\delta w + 3 \delta p \nonumber \\ \nonumber \\
&  & \partial_{j}{\dot{h^{j}_{i}}}-\partial_{i}{\dot{h}}-2 \dot{\beta_{it}}\partial_{j}{h^{j}_{t}}+2 \dot{\beta_{jt}}\partial_{j}{h^{t}_{i}}+\dot{\beta_{ij}}\partial_{j}{h}-\dot{\beta_{ab}}\partial_{i}{h^{a}_{b}}=-\frac{2}{\sqrt{\gamma}}(\omega + p) \delta u^{i} \nonumber\\ \nonumber \\
&  & \frac{1}{2}e^{-2\alpha}(e^{-2\beta})_{is}(\partial_{ia}{h^{a}_{s}}+\partial_{sa}{h^{a}_{j}}-(e^{2\beta})_{sk}(e^{-2\beta})_{ab}\partial_{ab}{h^{k}_{j}}-\partial_{sj}{h})  \nonumber \\ 
&+ & \frac{1}{4}\dot{\gamma}\dot{h^{i}_{j}}+\frac{1}{2}\gamma \ddot{h^{i}_{j}}+\frac{1}{2}\gamma \dot{h}(\dot{\alpha}\delta^{i}_{j}+\dot{\beta_{ij}})+\frac{3}{2}\gamma \dot{\alpha}\dot{h^{i}_{j}}+\gamma \dot{\beta_{is}}\dot{h^{s}_{j}}-\gamma \dot{\beta_{kj}}\dot{h^{i}_{k}}  
= \frac{1}{2}(\delta \omega-\delta p)\delta^{i}_{j} \nonumber \\
\end{eqnarray}
where $h \equiv\sum_{i=1,2,3}h^{i}_{i}$. 

Because spatial translation invariance is still a symmetry of the Bianchi I spaces, the modes can be separated by Fourier components. 
Let's select a particular wave-vector $\vec{k}$, then 

\begin{eqnarray}
h^{i}_{j}(\tau,\vec{x}) &  = & \mu^{i}_{j}(\tau)e^{i\vec{k}\cdot\vec{x}} \nonumber \\
\delta\omega & = & W(\tau)e^{i\vec{k}\cdot\vec{x}}  \nonumber \\
\delta p & = &  P(\tau)e^{i\vec{k}\cdot\vec{x}} \nonumber \\
\delta u^{i} & = & i V_{i}(\tau) e^{i\vec{k}\cdot\vec{x}} \nonumber \\ \label{FT}
\end{eqnarray}
and $\mu \equiv\sum_{i=1,2,3}\mu^{i}_{i}$. It is useful to introduce a projection operator $$\kappa_{ij}\equiv \delta_{ij}-\frac{k_{i}k_{j}}{k_{s}k_{s}}$$
It satisfies:  $\kappa_{is}\kappa_{sj}=\kappa_{ij}=\kappa_{ji}$, $\kappa_{is}k_{s}=0$ and $\kappa_{is}l_{s}=l_{i}$, where $\vec{l}\cdot\vec{k}=0$.\\

We define the following parts  of $\mu^{i}_{j}$

\begin{eqnarray}  \label{mueq}
\mu & \equiv & \sum_{s=1}^{3}\mu^{s}_{s}\\
r\ & \equiv  & \frac{1}{2}\left(\mu-\mu^{st}\frac{k_{s}k_{t}}{k_{a}k_{a}}\right)\\
q_{i} & \equiv & 2\mu^{s t}\left(\frac{k_{s}\kappa_{it}}{k_{a}k_{b}e^{-2\alpha}(e^{-2\beta})^{ab}}\right)\\
\eta^{i}_{j} & \equiv & \mu^{s t}\left(\kappa_{is}\kappa_{tj}-\frac{1}{2}\kappa_{ts}\kappa_{ij}\right) \label{mueq-1}.
\end{eqnarray}

For simplycity we will consider a single component for the wave vector, $\vec{k}=(0,0,k_{3})$. After substituting (\ref{mueq})-(\ref{mueq-1}) in (\ref{eq})
and doing some algebra, the equations (\ref{1}) and (\ref{2}) can be easily derived provided we make the identification:

$$ \delta \equiv \frac{W(\tau)}{w(\tau)} \hspace{7ex}  \mbox{and} \hspace{7ex} \eta \equiv  {\eta^{1}}_1$$ where $w(\tau) =T_{00}$. 

In the long wavelength limit with no residual isotropy ($s_{3}\neq \pm 2$), the equations (\ref{1}) and (\ref{2}) can be combined into a fourth-order differential equation for $\delta$, which can be solved to give $$\delta=\frac{A}{t}+B+Ct^{\frac{2(2-s_3)}{3}}\left(1+D \ln \left(\frac{t}{t_b} \right) \right)$$ Substituting this result in the equation for $\eta$, we have ($t \ll t_b$)

\begin{equation}
\frac{d^2 \eta}{dt^2}+\frac{1}{t}\frac{d \eta}{dt}+K^2t^{\frac{-2(1+s_3)}{3}}t_b^{\frac{-2(1-s_3)}{3}}\eta=\frac{1}{3}(s_1-s_2)\frac{1}{t}\frac{d\delta}{dt}
\label{as}
\end{equation}

For different choices of the parameters $s_i$ and of the coefficients in (\ref{5}), the behaviour of $\eta$ will be either increasing or decreasing as a function of time, but never oscillatory

\begin{equation}
\eta=C_1+\frac{C_2}{t}+C_3t^{\frac{2(2-s_3)}{3}}( 1 +C_4 \ln t)
\end{equation}
The constants $C_i$ are related to the constants that appear in (\ref{as}). 

The short wavelength limit translates into the condition $FK^2t^2 \gg 1$.
The trick to solve the equations in this limit is to average both $\delta$ and $\eta$ over an interval of time that needs to be relatively short in order to have a good approximation and, at the same time, long enough as to include several wavelengths.\\
The behaviour in the short wavelength approximation in an isotropic model (where the tensor and the scalar modes are decoupled), is oscillatory with a decreasing amplitude (see Lifshitz and Khalatnikov, 1963, page 513). 
In the anisotropic case the situation is different because tensor and scalar modes are coupled, nevertheless an oscillating behaviour is expected in the very short wavelength limit, with a matter density perturbation smoother than the tensor mode. An average over several wavelength for $\eta$ would thus be negligible compared to an average for $\delta$ if the former is represented as an oscillating function and the latter as a smoothly varying function.\\
With this is mind, dividing both sides of the equation (\ref{1}) by $FK^2t^2$ and averaging, for $t \ll t_b$ it simplifies as follows

\begin{equation}
\frac{d^2 \delta}{dt}+\frac{2(1+s_3)}{3t}\frac{d\delta}{dt}=0,
\end{equation} 
where now $\delta$ is an average quantity.\\
The solution to this equation is

\begin{equation}\label{31}
\delta=At^{\frac{(1-2s_3)}{3}}+B.
\end{equation}

Notice that the density contrast has, on top of an overall oscillating behaviour which has been subtracted thanks to the averaging procedure, a power law changing amplitude which is positive for a large part of the range of possible values of the parameter $s_3$.\\
Replacing this into the equation for $\eta$, we have

\begin{equation}\label{30}
\frac{d^2\eta}{dt^2}+\frac{1}{t}\frac{d \eta}{dt}+K^2F \eta=\frac{s_1-s_2}{3}A't^{-0.51},
\end{equation}
where the constant $A'$ turns out to be of the order of $10^{60}$ in units of $\mbox{sec}^{-1.49}$ according to our estimate of the parameters involved in perturbation theory, and we have assumed ,  $s_3=-\sqrt{3}$.
With a change of variable, we have 

\begin{equation}
\ddot{\eta}+\frac{\dot{\eta}}{x}+10^{4}x^{0.49}\eta=10^{-n}x^{-0.51}, 
\end{equation}
where the dots indicate the derivative with respect to $x \equiv \frac{t}{t_P}$, $t_P$ being the Plank time. The solution is a linear combination of Gamma, Hypergeometric and Bessel functions. In a simplified version of the previous equation setting the right hand side to zero (n is always bigger than one, so this is a reasonable approximation), the solution becomes

\begin{equation}\label{24}
\eta=C_1 J_{0}\left[4\sqrt{\frac{2}{5}}x^{\frac{5}{4}}\right]+C_2 Y_{0}\left[4\sqrt{\frac{2}{5}}x^{\frac{5}{4}}\right],
\end{equation}
with $C_1$ and $C_2$ constants of integration and $J_{0}$ and $Y_{0}$ are Bessel functions. This solution represents a sound wave with an amplitude that attenuates in time.\\
Since the rigth hand side of equation (\ref{30}) can be neglected without significantly altering the solution to the equation for $\eta$, we conclude that the choice of the anisotropy parameters $s_i$ doesn't really affect the overall behaviour of the tensor mode. This is radically different from the scalar mode, which is dramatically affected by that choice (\ref{31}).\\

%

\end{document}